\documentclass[journal,10pt]{IEEEtran}
\usepackage{cite}
\usepackage{color, xcolor}
\usepackage{graphicx}
\usepackage{subfigure}
\usepackage{amsmath}
\usepackage{amssymb}
\usepackage{amsfonts}
\usepackage{array}
\usepackage{float}
\usepackage{diagbox}
\usepackage{multirow}

\begin{document}
\title{ICI-Free Channel Estimation and Wireless Gesture Recognition Based on Cellular Signals}
\author{Rui Peng, \IEEEmembership{Student Member,~IEEE}, Yafei Tian, \IEEEmembership{Member,~IEEE}, and Shengqian Han, \IEEEmembership{Member,~IEEE}\\

\thanks{This work was supported by the National Natural Science Foundation of China under Grants 61971023.}
\thanks{The authors are with the School of Electronics and Information Engineering, Beihang University, Beijing 100191, P. R. China (e-mail: pengrui@buaa.edu.cn; ytian@buaa.edu.cn; sqhan@buaa.edu.cn).} }

\maketitle
\thispagestyle{empty}
\begin{abstract}
Device-free wireless sensing attracts enormous attentions since it senses the environment without additional devices. While cellular signals are good opportunistic radio sources, the influence of inter-cell interference (ICI) on wireless sensing has not been adequately addressed. In this letter, we first investigate the cause of ICI and its impact on wireless sensing. Then we propose an ICI-free channel estimation method by reconstructing the broadcast signals of adjacent cells and solving simultaneous equations. Wireless gesture recognition can be greatly benefited by ICI mitigation. Finally, we build a prototype system to receive the commercial 4G-LTE signals, and demonstrate the accuracies of wireless gesture recognition under various conditions. 
\end{abstract}

\begin{keywords}
Device-free wireless sensing, cellular signals, ICI, broadcast signals, gesture recognition.
\end{keywords}

\section{Introduction}\label{sec:introduction}
Integrated sensing and communication (ISAC) is a promising technology in the sixth generation (6G) mobile communication system. Device-free wireless sensing (DFWS) is one of the popular research areas in ISAC. Except for regular data transmission, the communication system can also help to identify human behaviors by capturing the variation of channel environment. The major benefit of DFWS is that it optimizes the utilization of time-frequency resources by integrating sensing tasks with communications, without any additional devices \cite{Paradigm}.

A majority of DFWS researches exploit WiFi as radio source, since it is convenient to extract the channel state information (CSI) by off-the-shelf network interface card. On the contrary, currently, there is no CSI extraction tool available for 4G/5G modem, so the acquisition of CSI relies on self-developed software defined radio (SDR). From the opportunistic signal perspective, cellular signal has broad coverage and continuous samplings of the CSI \cite{WIFI_Cellular}, making it a better choice for DFWS.

Due to the limitation of CSI accessibility, only a few works studied DFWS with 4G long term evolution (LTE) signal. In the literature, \cite{LTE_Crowd} proposed a crowd density estimation method through CSI, and proved that CSI can achieve better performance than reference signal received power (RSRP). The respiration detection was studied in \cite{LTE_res}. By utilizing the coverage advantage of LTE and combining the results measured in different frequency bands, the detection error is decreased and the blind spot issue is solved. The gesture recognition was explored in \cite{Robust}, where the direction of base station (BS) is first probed by a preamble gesture, and then the user stands in the opposite direction of the BS to improve the recognition performance. 

Despite these works achieved wireless sensing with LTE signals, none of them has investigated the influence of inter-cell interference (ICI). However, ICI is a practical problem in cellular networks and has critical impact on wireless sensing. Since the soft frequency reuse scheme is widely applied in current networks, most areas are covered by multiple cells operated in the same frequency band. Usually, CSI is estimated by cell-specific reference signal (CRS). While the CRSs from adjacent cells are inserted on different subcarriers to avoid collision, the CRS from one cell may still be contaminated by the data transmitted from another cell. For fine sensing tasks such as gesture recognition, the motion information is extracted from the subtle variation of propagation channels, and is more vulnerable to ICI. The recognition performance will be dramatically degraded whenever the strength of ICI exceeds the reflection power of human hand.
	
In this letter, we propose an ICI-free channel estimation method for wireless gesture recognition based on the time division duplexing (TDD) LTE cellular signals, and validate its effectiveness by prototype experiments. The main contributions of this work are as follows:

\hangafter 1
\hangindent 2.35em
1) We analyze the source of ICI and its characteristics, and clarify its impact on CRS-based channel estimation and subsequent gesture recognitions. 

\hangafter 1
\hangindent 2.35em
2)  We propose to use the physical broadcast channel (PBCH) as reference signal, so that the local sequences of ICI are known, and the channel estimation is transformed from an interference-limited problem to a noise-limited problem. 

\hangafter 1
\hangindent 2.35em
3) We develop the multi-cell PBCH local sequences reconstruction and joint channel estimation methods, and the ICI-free CSI can thus be acquired.

\hangafter 1
\hangindent 2.35em
4) We build a prototype system to receive commercial TDD-LTE signals and implement gesture recognition in real-time. Experiment results exhibit superior performance than existing methods.

\setlength \arraycolsep{1pt}
\section{System Model}\label{sec:system model}

\subsection{LTE background} \label{sec:LTE frame}
In LTE downlink (DL) signal, a system frame with duration 10ms is divided into  10 subframes. Each subframe comprises of 14 orthogonal frequency division multiplexing (OFDM) symbols in normal cyclic prefix (CP) case. A resource block (RB) is the basic scheduling unit, which consists of 12 subcarriers and each subcarrier in one symbol is called a resource element (RE). The CRS is inserted in each RB for channel estimation. The symbol index of CRS is fixed, but the subcarrier offset inside each RB is determined by physical layer cell identity (PCI), as
\begin{eqnarray}\label{eq:v_shift}
	k_0 =(v+v_\textrm{shift}) ~\textrm{mod} ~6, 
\end{eqnarray}
where $v$ is either 0 or 3 according to the antenna port and symbol number, $v_\textrm{shift} = N_\textrm{ID} ~\textrm{mod}~ 6$, and $N_\textrm{ID}$ denotes PCI. As a principle, the PCIs with different mod-6 value are assigned to adjacent cells, so that their CRSs will not collide.

In every DL frame, PBCH are broadcasted from symbol 7 to 10 in subframe 0. It occupies the rest resources other than CRS in the middle 72 subcarriers. PBCH carries master information block (MIB), which includes necessary system information such as system frame number (SFN) and bandwidth. The SFN is circularly updated from 0 to 1023, while the rest information bits remain constant. In addition, we can also implicitly obtain the antenna port number during PBCH decoding.

TDD systems separates uplink (UL) and downlink in the time domain. Although there are seven UL-DL subframe assignment configurations, subframe 0 and 5 are always for DL transmission\cite{36211_sim}.
To prevent UL-DL interference, the timing of adjacent TDD cells are well-synchronized, and they share the same UL-DL configuration and SFN. 

\subsection{Channel Model}\label{sec:channel model}
We consider a time-variant wideband channel. The CSI of a single stream, i.e., from one transmit port to one receive port, consists of static paths and dynamic paths. For symbol $n$ and subcarrier $k$, the CSI is denoted as
  \begin{eqnarray}\label{eq:baseband CSI}
	{h}(k,n)&=&e^{-j\theta_\textrm{p}(n)}\left[{h}_{\textrm{s}}(k) + {h}_{\textrm{d}}(k,n)\right],
\end{eqnarray}
where ${h}_{\textrm{s}}(k)$ is the static channel response, caused by LoS path and other paths reflected from static objects, ${h}_{\textrm{d}}(k,n)$ denotes the time-variant dynamic channel response induced by moving objects. 
However, the CSI is always polluted by the phase noise $\theta_\textrm{p}(n)$, which is introduced by oscillator vibration. Even with accurate frequency-offset tracking and compensation, the phase noise cannot be totally eliminated. 

\subsection{Impact of ICI} \label{sec:ICI}
\begin{figure}[htp!]
	\begin{center}
		\includegraphics[width=0.45\textwidth]{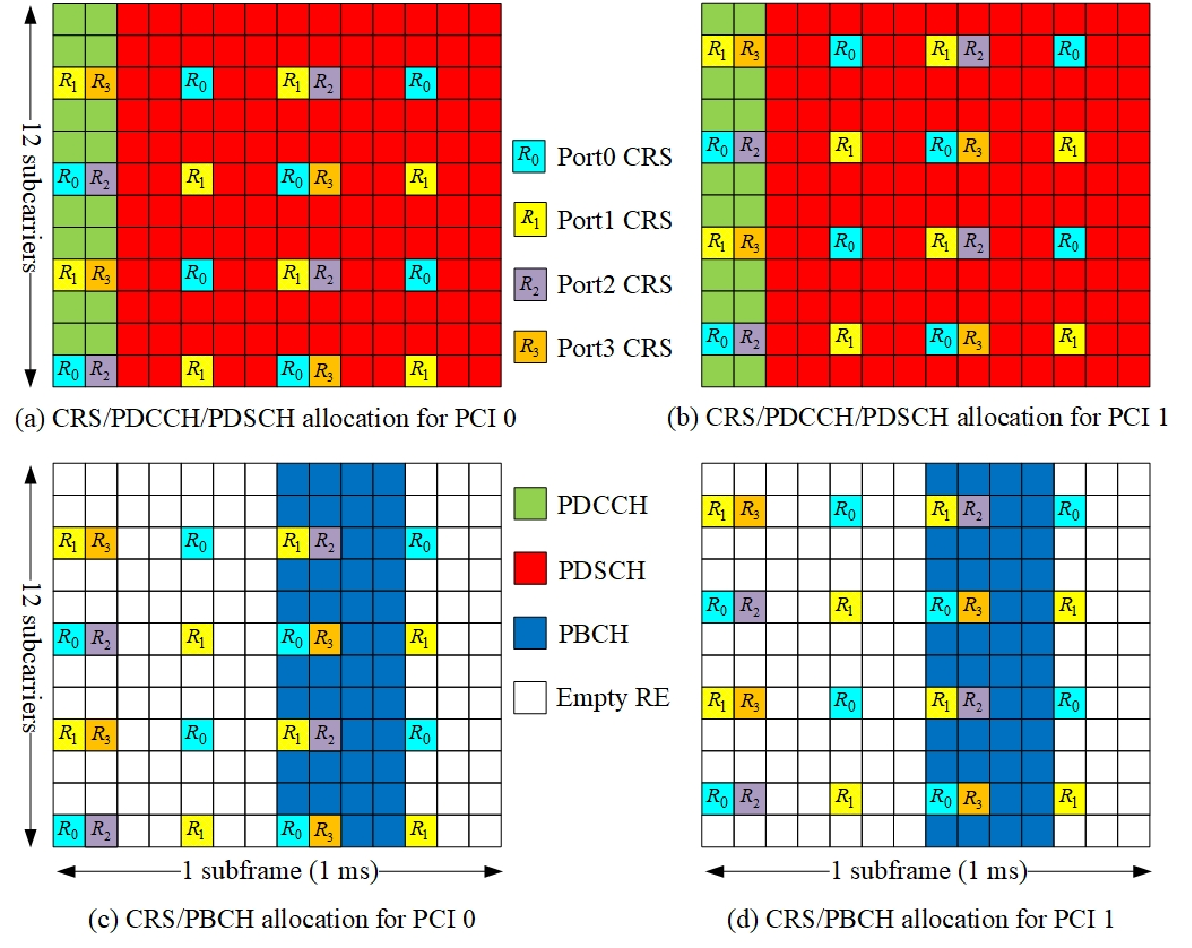}
		\centering \caption{An example of CRS/PDCCH/PDSCH/PBCH allocation for different cells in case of 4 antenna ports, normal CP.  The cases shown in figures (a)(b) and (c)(d) are in different RBs. }\label{fig.ICI}
	\end{center}
\end{figure}

In urban area, the cells are small and densely covered, so that users receive the superimposed signals from multiple cells. Usually, the cell with the highest RSRP is chosen as the serving cell. The adjacent cells will reuse the same time and frequency resources to improve the spectrum efficiency, thus introduces ICI.  Although CRSs from different cells will not collide,  they may collide with the physical downlink control channel (PDCCH) and physical downlink shared channel (PDSCH) of neighbor cells, as shown in Fig.\ref{fig.ICI}(a) and (b). 

As a result, the received CRS will be interfered as 
\begin{eqnarray}\label{eq:received CRS}
	y_{\textrm{CRS}}(k,n) &=& x_{\textrm{CRS}}(k,n)h(k,n) + \nonumber\\ &&\sum_{p=0}^{N_\textrm{I}-1}x^{p}_{\textrm{I}}(k,n)h^{p}_{\textrm{I}}(k,n) + w(k,n),
\end{eqnarray}
where $x_{\textrm{CRS}}(k,n)$ and $x^{p}_{\textrm{I}}(k,n)$ denote the sequence of CRS in serving cell and PDCCH/PDSCH payload in neighbor cell, respectively, $p$ is the index of interference data streams. We assume that there are $N_\textrm{I}$ interference streams in total. If all neighbor cells configure with one antenna port, $N_\textrm{I}$ is equal to the number of neighbor cells, $h(k,n)$ and $h^{p}_{\textrm{I}}(k,n)$ represent the frequency domain channel of the serving cell and the $p^\textrm{th}$ interference stream, respectively, and ${w}(k,n)$ is the additive white Gaussian noise. 

The CRS is an always-on signal, while the PDCCH/PDSCH is transmitted when the user equipment (UE) requests downlink data. In fact, the allocation of PDCCH/PDSCH is determined by BSs, and it is hard to decode the PDCCH/PDSCH of neighbor cells in the receiver, since the SINR is usually below zero.  

Consequently, the ICIs on CRS subcarriers are random and time-variant, and thus they are unknown and unpredictable. In the channel estimation results, they behave as parts of the dynamic channel, i.e.,
\begin{eqnarray}\label{eq:CSI CRS}
	\hat{h}(k,n) = h(k,n) +  \frac{\sum_{p=0}^{N_\textrm{I}-1}x^{p}_{\textrm{I}}(k,n)h^{p}_{\textrm{I}}(k,n) + w(k,n)}{x_{\textrm{CRS}}(k,n)}.
\end{eqnarray}
Although the UE typically accesses the strongest cell, resulting in a potentially weaker interference compared to the channel response of the serving cell, the interference term can still be much stronger than the reflection path of human hand, causing severe impact on dynamic CSI extraction and subsequent gesture recognition.

\section{Methodology}\label{sec:Methodology}
\subsection{Multi-Cell Joint Channel Estimation}\label{sec:PBCH Channel Estimation}
To eliminate the impact of ICI, we propose a multi-cell joint channel estimation method. Instead of using CRS, we use PBCH as reference signal.  The PBCH of serving cell and neighbor cells are always overlapped in the same $4\times72$ RE grid in subframe 0, as shown in Fig.\ref{fig.ICI}(c) and (d). Although the PBCHs from neighbor cells compose ICIs as well, the transmitted sequences of all PBCHs can be predicted, since the generation of PBCH follows specific rules. By solving simultaneous equations about the received signals and known local sequences, we can acquire ICI-free channel estimation.

\subsubsection{Neighbor cells search and PBCH reconstruction}
As introduced in Section \ref{sec:LTE frame}, PBCH carries MIB, wherein the only changed information is the SFN bits. Due to the frame transmission mode in LTE, the SFN bits in all frames can be derived by the SFN known in one frame. Once we decode the MIB in any frame, the PBCH sequences can be reconstructed in all frames. It means that we only need to decode the PBCHs for one time. Moreover, the PBCH can be decoded in low SINR, since the coding rate is 1/3 and the coded bits are repeated multiple times in rate matching. 
 
To correctly decode PBCHs of all adjacent cells, a successive interference cancellation (SIC) strategy is required. In TDD-LTE, not only the PBCH, but also the primary synchronization signal (PSS) and secondary synchronization signal (SSS) from different cells are all aligned. After cell search and frequency offset compensation using the PSS and SSS, the UE can decode MIB of the serving cell. To detect the neighbor cells, we need first reconstruct the PSS/SSS/PBCH signals of the serving cell, and cancel them from the corresponding REs. Then we can search and decode the next stronger neighbor cell. In case that we successfully decode the MIB of a new cell, we list it as a neighbor cell. The SIC procedures are repeated until no further cell can be found.

Despite the complexity of SIC is high, it is only conducted in the preparation stage, and it will not affect the real-time processing in channel estimation and gesture recognition. The reconstruction of PBCH consists of convolutional encoding, rate matching, scrambling, modulation and transmit diversity, as elaborated in 3GPP specifications 36.211 \cite{36211_sim} and 36.212 \cite{36212_sim}. Due to space limitations, we will not go into details here.  

\subsubsection{Joint channel estimation}\label{sec:Multi-stream CSI}
Considering a general scene, at one receiver, totally $N_\textrm{s}$ streams of PBCH signals are received from adjacent cells. In frame $n$, symbol $l$ and subcarrier $k$, the received signal at one RE can be represented as
\begin{eqnarray}\label{eq:PBCH_equation}
	y(k,l,n) = \sum_{p=0}^{N_\textrm{s}-1}x^{p}(k,l,n)h^{p}(k,l,n) + w(k,l,n),
\end{eqnarray}
where $x^{p}(k,l,n)$, $h^{p}(k,l,n)$ and $w(k,l,n)$ denote the PBCH modulated symbol, frequency domain channel response and noise, respectively.

In (\ref{eq:PBCH_equation}), although $x^{p}(k,l,n)$ is known after PBCH reconstruction, we have $N_\textrm{s}$ unknown variables $h^{p}(k,l,n)$ to be estimated in each equation. Fortunately, the channel of adjacent REs typically have only minor changes. We consider that $h^{p}(k,l,n)$ keeps constant in a RE group, which consists of $K$ contiguous subcarriers and $L$ symbols. To ensure that equation (\ref{eq:PBCH_equation}) has a unique solution, the selection of $K$ and $L$ should satisfy $KL\ge N_\textrm{s}$. 

Since PBCH occupies 72 subcarriers and 4 symbols in a frame, it has plenty of freedom to separate RE groups. We can slightly increase the number of $KL$, so that the equations are overdetermined and the estimation error can be decreased by least square (LS) method.

With the assumption that channel is invariant inside a RE group $[k_\textrm{g}, l_\textrm{g}]$, (\ref{eq:PBCH_equation}) can be approximated as 
\begin{eqnarray}\label{eq:PBCH_matrix}
      	y(k,l,n) \approx \sum_{p=0}^{N_\textrm{s}-1}x^{p}(k,l,n)h^{p}_{k_\textrm{g},l_\textrm{g}}(n) + w(k,l,n) \nonumber\\ {\forall}k,l \in \mathbb{I}_{k_\textrm{g},l_\textrm{g}},
\end{eqnarray}
where $h^{p}_{k_\textrm{g},l_\textrm{g}}(n)$ is the channel to be solved for RE group $[k_\textrm{g}, l_\textrm{g}]$, and
 $\mathbb{I}_{k_\textrm{g},l_\textrm{g}}$ denotes the subcarrier and symbol index set in RE group $[k_\textrm{g}, l_\textrm{g}]$. 
Write (\ref{eq:PBCH_matrix}) in matrix form, 
\begin{eqnarray}\label{eq:PBCH_matrix_abbr}
	\mathbf{y}_{k_\textrm{g}, l_\textrm{g}}(n) \approx \mathbf{X}_{k_\textrm{g}, l_\textrm{g}}(n)\mathbf{h}_{k_\textrm{g}, l_\textrm{g}}(n) + \mathbf{w}_{k_\textrm{g}, l_\textrm{g}}(n),
\end{eqnarray}
and we can obtain the LS estimation of $\mathbf{h}_{k_\textrm{g}, l_\textrm{g}}(n)$ as
\begin{eqnarray}\label{eq:PBCH_LS}
	\hat{\mathbf{h}}_{k_\textrm{g}, l_\textrm{g}}(n) = [\mathbf{X}^H_{k_\textrm{g}, l_\textrm{g}}(n)\mathbf{X}_{k_\textrm{g}, l_\textrm{g}}(n)]^{-1}\mathbf{X}^H_{k_\textrm{g}, l_\textrm{g}}(n)\mathbf{y}_{k_\textrm{g}, l_\textrm{g}}(n).
\end{eqnarray}

It can be seen that this is a joint channel estimation method by utilizing the PBCH signals of all adjacent cells. Except some estimation errors and approximation errors, there is no longer influence of ICI. For simplicity, we only give LS estimation here, actually more sophisticated estimation method can be developed based on this idea, such as exploiting the channel correlation in time and frequency domains. Furthermore, since the sensing unit just listens the  opportunistic signals on air and do not affect the transmit procedure in BSs, there is no additional backhaul or signaling consumption.

The simultaneous equations can also be constructed by PSS/SSS, but PSS/SSS occupy less REs and only support single stream transmission. Furthermore, this method can also be used in 5G new radio (NR), where the synchronization signal and PBCH block (SSB) have similar characteristics as the PBCH in LTE.

\subsection{CSI Processing and Gesture Recognition}\label{sec:CSI processing}
After channel estimation, the gestures can be recognized by analyzing the CSI variation of the serving cell.
We combine the CSI in multiple subcarriers by inverse Fourier transform (IFFT), and extract the strongest path, which can significantly improve the signal to noise ratio (SNR) of the channel estimation results\cite{multiuser_TVT}.
However, as shown in (\ref{eq:baseband CSI}), even without ICI the CSI is also impacted by phase noise. CSI ratio is an effective way to eliminate the phase noise \cite{FarSense}. If the serving cell is equipped with two or more antenna ports, we can select one port as the reference channel. Otherwise, we use two receive antennas and set one as the reference.

For PBCH-based channel estimation, we can update CSI every 10ms, which corresponds to a sampling frequency of the consecutive CSI at 100Hz. According to Nyquist sampling theory, the maximum Doppler frequency that can be identified is 50Hz. Fortunately, most commercial LTE bands operate at frequencies below 3GHz. Therefore, even if a hand moves at a speed of 1m/s,  the Doppler shift will not exceed 20Hz. To extract the dynamic path $\hat{h}_\textrm{d}(n)$, a band pass filter is used to separate the channel response with frequency shift range of 3-20Hz from the denoised channel.

\begin{figure}[htp!]
	\begin{center}
		\includegraphics[width=0.4\textwidth]{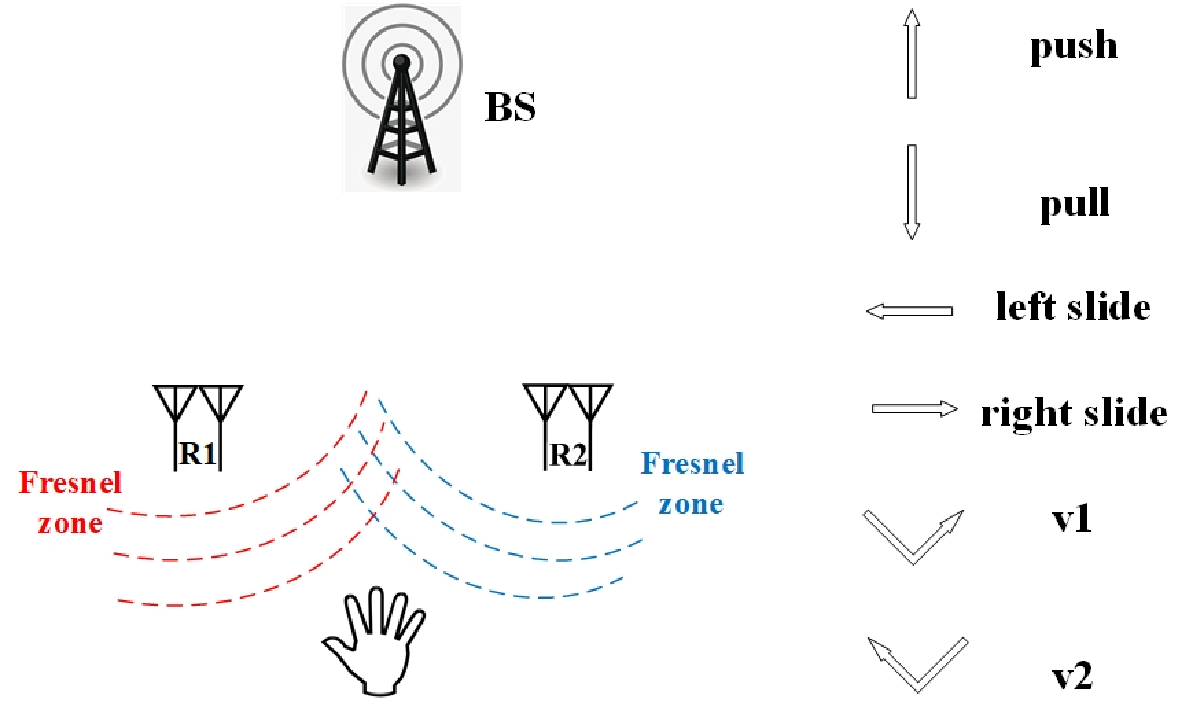}
		\centering \caption{Candidate gestures and user position.}\label{fig.gesture}
	\end{center}
\end{figure}

In this work, we utilize two receivers to recognize the two-dimensional gestures. As shown in Fig. \ref{fig.gesture}, totally 6 candidate gestures are designed, and the user locates in the central area between two receivers and faces to the BS. The direction of the BS can be identified through the solution proposed in \cite{Robust}.

Since the focus of this letter is on channel estimation and the gesture patterns are simple, we use a training-free method to recognize different gestures based on the change rule of Doppler shifts. We employ a sliding window with duration $K_\textrm{D}$ to calculate the Doppler spectrogram by short-time discrete Fourier transform (DFT).  In the period of gesture performing, we identify the frequency component with the highest amplitude in each sliding window, and form the Doppler shift vector in two receivers, represented by $\mathbf{f}_1$ and $\mathbf{f}_2$, respectively. 

After removing the outliers in $\mathbf{f}_1$ and $\mathbf{f}_2$, we first check whether there is a short pause in the motion according to the values of $|\mathbf{f}_1| + |\mathbf{f}_2|$, where `$|\cdot|$' is the element-wise absolute operator. The gestures `v1/v2' can be distinguished if it is true, and otherwise the gesture should belongs to the other four possibilities. In Fig. \ref{fig.gesture}, it can be observed that a push or pull movement results in positive or negative Doppler shifts in both R1 and R2, so the gesture will be identified as a push or pull when all the non-zero elements in both $\mathbf{f}_1$ and $\mathbf{f}_2$ have the same signs. For the movement of left or right slide,  the Doppler shifts of two receivers are in the opposite direction. Therefore, we can differentiate the left or right slide by the sign of the accumulated value of  $(\mathbf{f}_1 - \mathbf{f}_2)$. The same criterion is applied to differentiate `v1' and `v2'.

\section{Evaluations }\label{sec:Experiments}
\subsection{Experiment Environment and Configuration}\label{sec:Environment}
\begin{figure}[htp!]
	\begin{center}
		\includegraphics[width=0.4\textwidth]{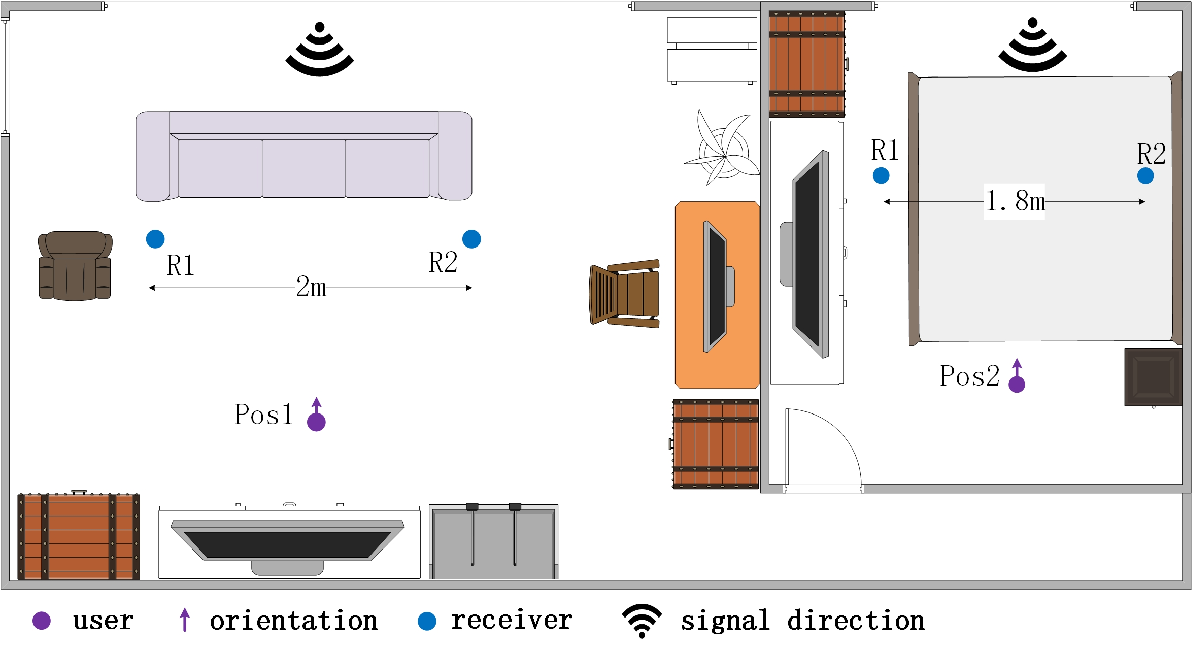}
		\centering \caption{The layout of the experiment environment. }\label{fig.home}
	\end{center}
\end{figure}

To evaluate the effectiveness of the proposed method, we conduct some experiments in an apartment using commercial LTE signals. The layout of the rooms are shown in Fig. \ref{fig.home}, and the experiments are implemented in the living room and bedroom. The commercial signals are from China Mobile TDD-LTE network operated in Band 40, centering in 2.3498GHz with 20MHz bandwidth. Both rooms share the same serving cell with PCI 252, and there are also three neighbor cells with PCI 249, 253, 256. From the PBCH decoding result, we find that they all utilize only one antenna port. We set the parameters of joint channel estimation, $K$ and $L$, as 3 and 4, respectively. The signals are transmitted by outdoor macro-BSs from hundreds meters away. There are no line-of-sight paths from the receiver to all BSs, but we observe that the signal of serving cell is mostly penetrated from the window.

We design a prototype system to achieve real-time channel estimation and gesture recognition. The RF signal is received and sampled by a software radio platform YunSDR Y550s, which equips 4 antennas and supports 4 streams of data receiving with maximal bandwidth of 100MHz. The baseband processings and gesture recognition are implemented in the host computer with an Intel Core i7-8700 CPU working at 3.20GHz. We place two antennas at position R1 and the other two at position R2, as if we have two receivers. Employing two antennas at one position is to eliminate the influence of phase noise.  

\subsection{Experiment Results}\label{sec:Result}
In the experiments, we start with measuring the signal strength of all cells using RSRP, which is defined as the average power of CRS. On one hand, RSRP indicates the signal quality of the serving cell. On the other hand, the RSRP of neighbor cells can also reflect the strength of ICI, since in LTE the transmission powers of PDCCH and PDSCH have fixed ratio with that of CRS. The RSRP results are shown in Table \ref{T:RSRP}. Due to the indoor multipath propagation, the RSRP fluctuates across different test positions. But it can be seen that the ICI is strong.  

\begin{table}
	\caption{MEASURED RSRP RESULTS}\label{T:RSRP}
	\centering
	\begin{tabular}{c|c|c|c|c}
		\hline
		\multirow{2}{*}{\diagbox{$N_\textrm{ID}$}{Rx}}	& \multicolumn{2}{c|}{Living Room}  & \multicolumn{2}{c}{Bedroom}  
		\\ \cline{2-5} &  R1 & R2 & R1 & R2 
		\\\hline
		252 & -77dBm & -71dBm & -82dBm & -76dBm\\ \hline
		249 & -86dBm & -85dBm & -101dBm & -88dBm \\\hline
		253 & -85dBm & -81dBm & -92dBm & -86dBm \\ \hline
		256 & -100dBm & -94dBm & -95dBm & -93dBm \\\hline
	\end{tabular}
\end{table}

\begin{figure}[htp!]
\begin{center}
	\includegraphics[width=0.48\textwidth]{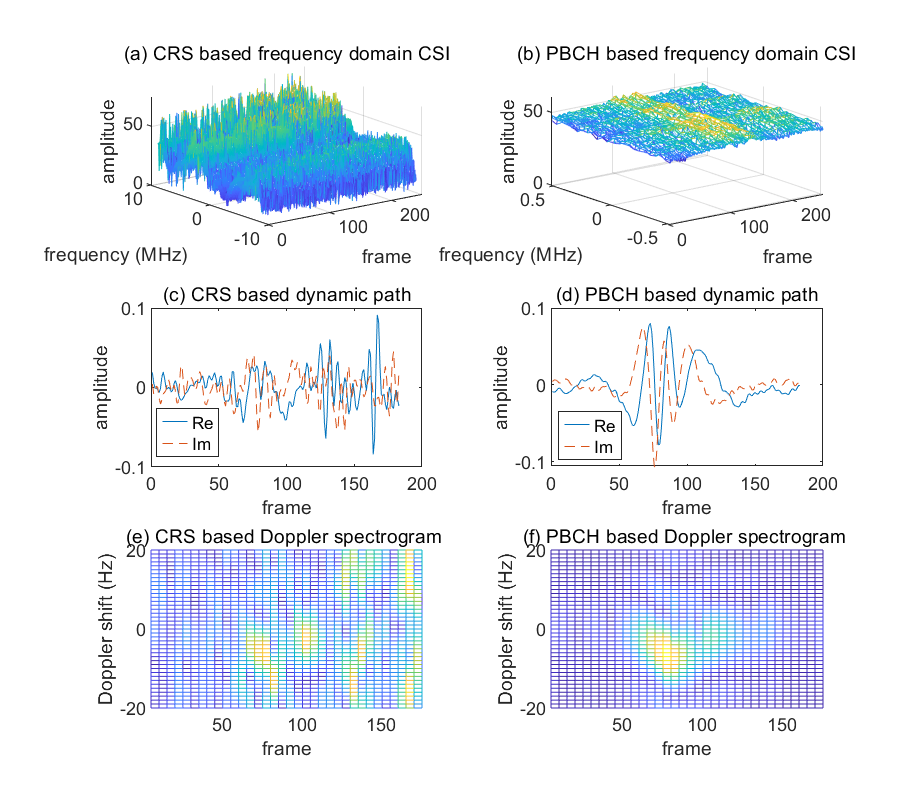}
	\centering \caption{An example of CRS-based and PBCH-based channel estimation and gesture recognition. (a) The CRS-based channel estimation results in full bandwidth. (b) The PBCH-based channel estimation results in the middle 72 subcarriers. (c-d) The CRS-based/PBCH-based dynamic path extraction. (e-f) The Doppler spectrogram calculated from CRS-based/PBCH-based channel estimation.}\label{fig.example}
\end{center}
\end{figure}

To illustrate the impact of ICI, we present an example in Fig. \ref{fig.example}, where a pull gesture is performed. As shown in Fig. \ref{fig.example}(a), if we obtain the CSI of serving cell by CRS, we observe numerous spurs in frequency domain. After main path extraction, denoising and filtering, the gesture pattern is still obscured by the ICI, as depicted in Fig. \ref{fig.example}(c). Consequently, the Doppler spectrogram is messy as illustrated in Fig. \ref{fig.example}(e). On the contrary, for the CSI acquired by PBCH-based joint channel estimation, the frequency domain channel response is smooth, the extracted dynamic path waveform and the corresponding Doppler spectrogram are clear and evident, as can be seen in Fig. \ref{fig.example}(b)(d)(f), respectively. In Fig. \ref{fig.example}, the length of sliding window is 100ms, and we calculate the DFT at 1Hz interval.

Recognition accuracy is the most important performance indicator for wireless gesture recognition. Next, let us compare the recognition accuracies according to the CSIs acquired by the CRS-based and PBCH-based channel estimations, which are collected in the same groups of tests. For CRS-based CSI acquisition, we obtain the frequency domain CSI by LS estimation in each subframe 0. Furthermore, we consider a subcarrier selection scheme (CRS-SS) based on the raw CRS-based CSIs, which is proved to be effective in combating the co-channel interference in WiFi sensing. The subcarriers are chosen with the weakest correlations after dynamic time warping \cite{CCI_WIFI}.

\begin{figure}[htp!]
	\begin{center}
		\includegraphics[width=0.45\textwidth]{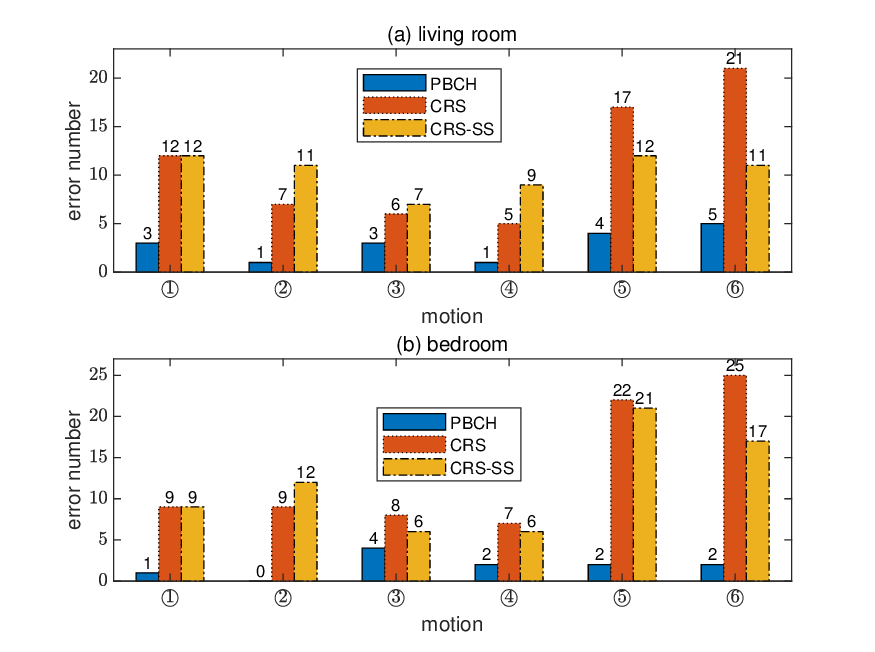}
		\centering \caption{A comparison of error number for each gesture between the PBCH-based, CRS-based and CRS-SS based approaches. Recognition accuracies: (a) Living room, PBCH $94.3\%$, CRS $77.3\%$, CRS-SS $79.3\%$; (b) Bedroom, PBCH $96.3\%$, CRS $73.3\%$, CRS-SS $76.3\%$.}\label{fig.accuracy}
	\end{center}
\end{figure}

In the test, we perform each type of gesture for 50 times. The error recognition, false alarm, and missed detection are all counted as errors. In CRS-based approach, the full 20MHz bandwidth is utilized, while the PBCH-based approach only utilizes 1.4MHz bandwidth it occupied. In the CRS-SS scheme, the highest accuracies in the living room and bedroom are achieved by selecting 100 and 120 subcarriers, respectively, out of the 200 CRS subcarriers. The error number statistics of each gesture is shown in Fig. \ref{fig.accuracy}. 

We can see that the proposed PBCH-based method comprehensively outperforms the other two methods. Although the CSI estimated by CRS can be smoothed by 200 subcarriers in 20MHz bandwidth, the ICI severely deteriorates the performance. With subcarrier selection, the accuracies show some improvement, but it is still far from the PBCH-based method in both scenarios. This is because the ICI typically spreads over the entire bandwidth, making it difficult to select clean subcarriers.

\section{Conclusion}\label{sec:conclusion}
This paper proposed an ICI-free channel estimation and wireless gesture recognition scheme by employing TDD-LTE signals. Firstly, we analyzed the cause of ICI and its impact on wireless sensing. Then, we proposed a multi-cell joint channel estimation method based on the reconstructed  broadcast signals from all adjacent cells. The proposed method can eliminate ICI with only one receive antenna and the minimum LTE system bandwidth. Finally, we developed a prototype system to receive commercial LTE signals and perform gesture recognition. The experiment results demonstrated that the CRS-based approach was susceptible to ICI, while the proposed method achieved high recognition accuracies.

\end{document}